\documentclass[smallextended,final]{svjour3}
\usepackage{epsfig}
\usepackage{amsmath}
\usepackage{mathptmx}

\newcommand{\ups}{\rule{0pt}{12pt}}

\journalname{Experimental Astronomy}

\begin{document}

\title{Robotic optical telescopes global network MASTER II.\\ Equipment, structure, algorithms.}

\author{Victor~Kornilov \and Vladimir~Lipunov \and Evgeny~Gorbovskoy \and Aleksander~Belinski \and
Dmitry~Kuvshinov \and Natalia~Tyurina \and Nikolai~Shatsky \and Anatoly~Sankovich \and Aleksander~Krylov \and Pavel~Balanutsa \and Vadim~Chazov \and Artem~Kuznetsov \and Dmitry~Zimnuhov \and Victor~Senik \and
Andrey~Tlatov \and Aleksander~Parkhomenko \and Denis~Dormidontov \and
Vadim~Krushinsky \and Ivan~Zalozhnyh \and Aleksander~Popov \and
Sergey~Yazev \and Nikolai~Budnev \and Kirill~Ivanov \and Evgeny~Konstantinov \and
Oleg~Gress \and Oleg~Chvalaev \and
Vladimir~Yurkov \and Yury~Sergienko \and Irina~Kudelina}

\institute{
V.G.\,Kornilov (\email{victor@sai.msu.ru}) \and V.M.\,Lipunov \and E.S.\,Gorbovskoy \and A.A.\,Belinski \and D.A.\,Kuvshinov \and N.V.\,Tyurina \and A.V.\,Sankovich \and A.V.\,Krylov \and N.I.\,Shatsky \and P.V.\,Balanutsa \and V.V.\,Chazov \and  A.S.\,Kuznetsov \and D.S.\,Zimnuhov \and V.A.\,Senik
\at Moscow state university, Sternberg astronomical institute, Moscow, Russia
\and
A.G.\,Tlatov \and A.V.\,Parkhomenko \and D.V.\,Dormidontov
\at High altitude astronomical station of the Pulkovo observatory, S-Peterburg, Russia
\and
V.V.\,Krushinsky \and I.S.\,Zalozhnyh \and A.V.\,Popov
\at Ural state university, Yekaterinburg, Russia
\and
S.A\,Yazev \and N.M.\,Budnev \and K.I.\,Ivanov \and E.N.\,Konstantinov \and O.A.\,Gress \and O.V.\,Chvalaev
\at Irkutsk state university, Irkutsk, Russia
\and
V.V.\,Yurkov \and Yu.P.\,Sergienko \and I.V.\,Kudelina
\at Blagoveshchensk pedagogical university, Blagoveshchensk, Russia
}

\date{Received: September 7, 2011 / Accepted: November 28, 2011}


\authorrunning{Kornilov et al}

\maketitle

\begin{abstract}
Presented paper describes the basic principles and features of the implementation of a robotic network of optical telescopes MASTER, designed to study the prompt (simultaneous with gamma radiation) optical emission of gamma-ray bursts and to perform the sky survey to detect unknown objects and transient phenomena. With joint efforts of Sternberg astronomical institute, High altitude astronomical station of the Pulkovo observatory, Ural state university, Irkutsk state university, Blagoveshchensk pedagogical university, the robotic telescopes MASTER~II near Kislovodsk, Yekaterinburg, Irkutsk and Blagoveshchensk were installed and tested. The network spread over the longitudes is greater than 6 hours. A further expansion of the network is considered.
\end{abstract}
\keywords{GRB  \and Optical observations \and Transients \and Supernovae \and Robotic observations}

\section{Introduction}

Since the beginning of the new century, it was realized that robotic observatories boost the capabilities of astronomical observations of non-stationary and short-lived phenomena in the Universe. Such facilities built throughout the world allowed to discover and study the prompt emission of the most powerful explosions --- the gamma-ray bursts. Hundreds of supernovae were detected as well which enabled to infer the existence of the so called ``dark energy'' or the energy of the cosmic vacuum. Finally, the robotised telescopes discover thousands of new minor bodies of the solar system and many exoplanets.

In order to serve efficiently these studies, the small apertures suite quite well \cite{Pach06}, but what is more important is the large field of view. Unpredictability of majority of transient phenomena and their practically isotropic distribution on the sky make probability of detection and measurement also proportional to the available yearly observation time at the site. Finally, the relative ease of installation of a small telescope makes possible to grow the overall efficiency by expansion of the observational network placing the new telescopes in various geographical points \cite{Bootes2010,PROMPT,ROTSE-III,TAROT}.

Russia, being the most longitude-extended country, is very attractive for studies of transients. The first Russian robot-telescope MASTER was developed in 2002--2006 and installed nearby Moscow \cite{AZh2007}. It appeared not only scientifically fruitful \cite{AZh2007,PAZh2008} but also served as an effective platform for methodology and technical studies devoted to optimise the use of small robotized telescopes of which the most essential is the unique software complex for automated multi-purpose astronomical observations.

The further development of the MASTER network was performed in 2008--2010 with deployment of four new observatories equipped with MASTER~II telescopes and very wide field (VWF) cameras \cite{Ada2010c,Ada2010a}. A key feature of this system is a small twin-tube telescope enabling measurement of brightness of transient phenomena simultaneously in two spectral bands or with two different directions of the axes of polarizing filters. Clearly, such an information is deadly needed for building adequate models of observed events. This paper describes the properties of the telescopes installation sites and provides the analysis of the global MASTER network efficiency as well as its individual nodes impact. Then the structure and functioning of the nodes basic soft and hardware complex is considered taking into account its centralized scheduling.

The MASTER~II telescopes network is involved in the following observational tasks:
\begin{itemize}
\item[---] Synchronous multi-colour and polarimetric observations of gamma-ray bursts;
\item[---] Supernovae search;
\item[---] Exoplanets observations;
\item[---] Trans-neptunian objects, comets and meteors detection;
\item[---] Study of orphan afterglows;
\item[---] Gravitational microlensing observation.
\end{itemize}

These observations are related to solving the following fundamental problems:
\begin{itemize}
\item[---] Dark energy study;
\item[---]  Gamma-ray bursts nature study;
\item[---]  Planet systems origin.
\end{itemize}

\section{MASTER network nodes location}

A small telescope, together with all its detectors and relevant equipment and maintenance, is incomparable to a large telescope by the financial resources involved. This is a reason why such projects like MASTER network cannot bear the full load of infrastructure support related to any modern observatory. Recalling also the insensitivity of the fast survey telescope optics to the atmospheric seeing, this effectively drives the selection of potential sites for installation.

While choosing the nodes location, we considered the following primary factors: 1) the longitude distribution of nodes; 2) the existence of the local personnel interested in joint development and scientific collaboration and sharing of a common infrastructure; 3) the clear skies night-time fraction. The experience and results of work of similar small-telescope programs  support the validity of these criteria (see e.g.  \cite{Pach06,PROMPT,rotse,Bootes}).

\begin{table}[b]
\caption{ Geographical coordinates of existing and potential MASTER network observatories.\label{tab:0}}
\smallskip
\centering
\begin{tabular}{lrrr}
\hline\hline
Node   & Longitude & Latitude & Altitude, m  \ups \\[4pt]
\hline
MASTER-URAL \ups  & $03^h58^m11^s\!.2$  & $+57^\circ 02^\prime 13^{\prime\prime}$ & 290 \\
MASTER-TUNKA      & $06^h52^m16^s\!.1$  & $+51^\circ 48^\prime 34^{\prime\prime}$ & 700 \\
MASTER-AMUR       & $08^h29^m56^s\!.0$  & $+50^\circ 19^\prime 07^{\prime\prime}$ & 215 \\
MASTER-KISLOVODSK & $02^h50^m04^s\!.0$  & $+43^\circ 45^\prime 00^{\prime\prime}$ & 2067 \\[4pt]
\hline
MASTER-CANARIES\ups & $01^h06^m02^s\!.3$  & $+28^\circ 18^\prime 00^{\prime\prime}$ & 2390 \\[3pt]
\hline\hline
\end{tabular}
\end{table}

Table~\ref{tab:0} provides the geographical data for four working and one under-development observatories of the network. The site MASTER-URAL is situated on the territory of Kourovskaya astronomical observatory of Ural state university since fall 2008. It is far enough ($\approx 80$~km) from the nearest city Yekaterinburg so the primary source of the moderate light pollution (see Table~\ref{tab:1}) is two railway stations 5~km away. This site is the most northern node of the network so some time in summer is characterised by the absence of astronomical night.

Installation of the MASTER-KISLOVODSK complex was also started in late 2008 with assembly of the clam-shell dome atop the 9-m high steel tower remained from solar astroclimate study program in 1970s. The tower is hosted by the solar station of the Pulkovo observatory which is 30~km to the south from Kislovodsk city. Note that developing Caucasus observatory of Sternberg astronomical institute (SAI) is located less than 1~km apart from the solar station \cite{kgo2010}. The southern neighbourhood of the site in direction to the Caucasus main ridge 40~km away is free from any light sources while the northern part of horizon is illuminated by towns of the Caucasus mineral water region.

In summer 2009 the temporary optical tube assembly (Flugge optical system, diameter 280~mm, field $3\times 3^\circ$) was replaced with a MASTER standard instrument (40~cm Hamilton system twin telescope) which enabled its full-featured operation. In parallel, two VWF installations are in operation at the site since 2006 (at the solar station as well) and 2008 (at Caucasus observatory of SAI, see \cite{Ada2010b}). This node is currently the most western and southern in the network, located also at the highest elevation above the sea level.

\begin{figure}
\centering
\psfig{figure=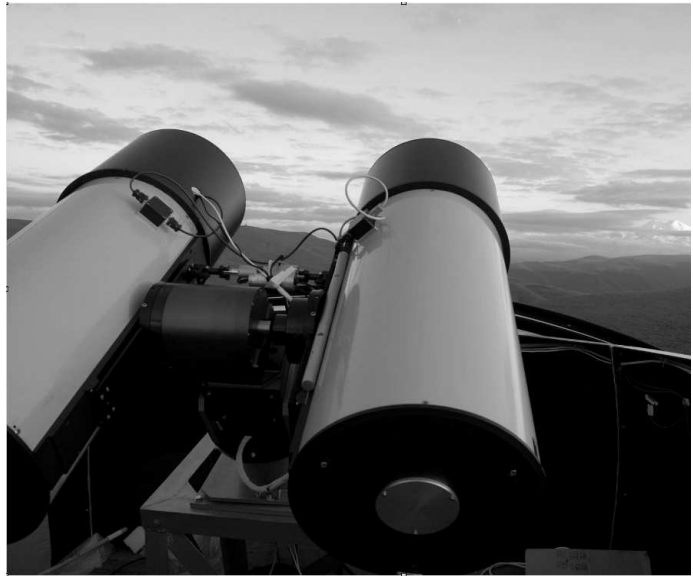,width=9cm}
\caption{The MASTER~II telescope at Pulkovo solar station (the MASTER-KISLOVODSK node of the network) on the top of 9~m steel tower. Black cylinder atop of each tube is a dew cap inside of which the photometric unit with a CCD camera is placed. \label{fig:m2-kgo}}
\end{figure}

The MASTER-TUNKA node is built in the Tunka valley (Buryat Republic of Russia) some 50~km from southern end of Baikal lake, at the proof ground of the Applied Physics institute of Irkutsk state university. The complex assembly was started in 2009 and the automated custom full-sky steel enclosure was then erected. In November 2009, this site entered into operation with the same temporary Flugge telescope equipped with Alta~16U CCD-camera.

The standard Hamilton twin telescope replaced the temporary tube in summer 2010 which allowed the site to reveal its great potential. The site benefits from its geographic position being distant from large towns and possessing high atmospheric transparency. Light pollution from the Irkutsk region cities is fully blocked by Eastern Sayan mountain ridge.

\begin{figure}
\centering
\psfig{figure=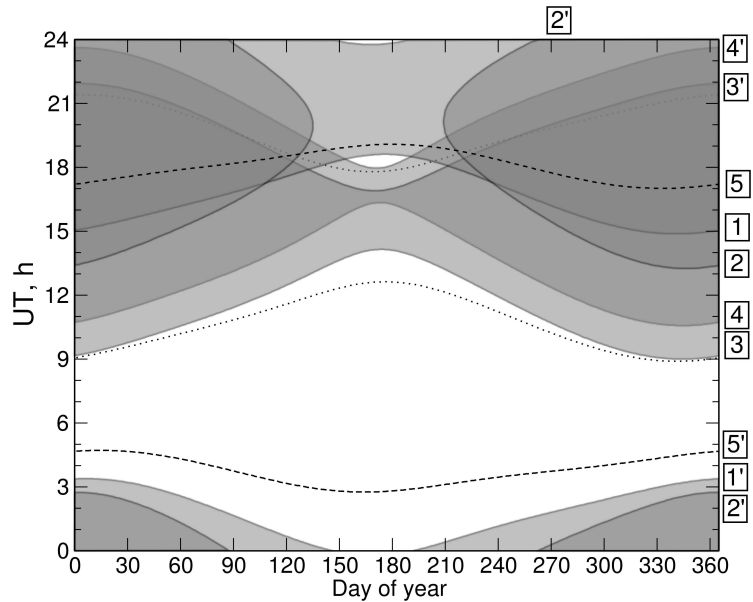,width=10cm}
\caption{Night-time duration diagram for various MASTER sites. Curves shown are: 1/1'  -- the night beginning/end at Kislovodsk; 2/2' -- the same for Kourovskaya observatory; 3/3' -- Blagoveschensk site; 4/4' -- Tunka proof ground; 5/5' -- planned MASTER~II installation at Tenerife island, Canaries. The dotted line depicts night range for Ussuriysk astronomical station.
\label{fig:time_coverage}}
\end{figure}

The eastern point of the network (MASTER-AMUR) is located at the territory of the discontinued Blagoveschensk latitude station two kilometers from Blagoveschensk city \cite{yurkov}. Alike other sites, half a year of construction and assembly works were completed in November 2009 with a temporary robot-telescope installation in the custom steel enclosure and the standard telescope tubes were put into operation a year later. The site operates full-time since January 2011.

\begin{figure}
\centering
\begin{tabular}{cc}
\psfig{figure=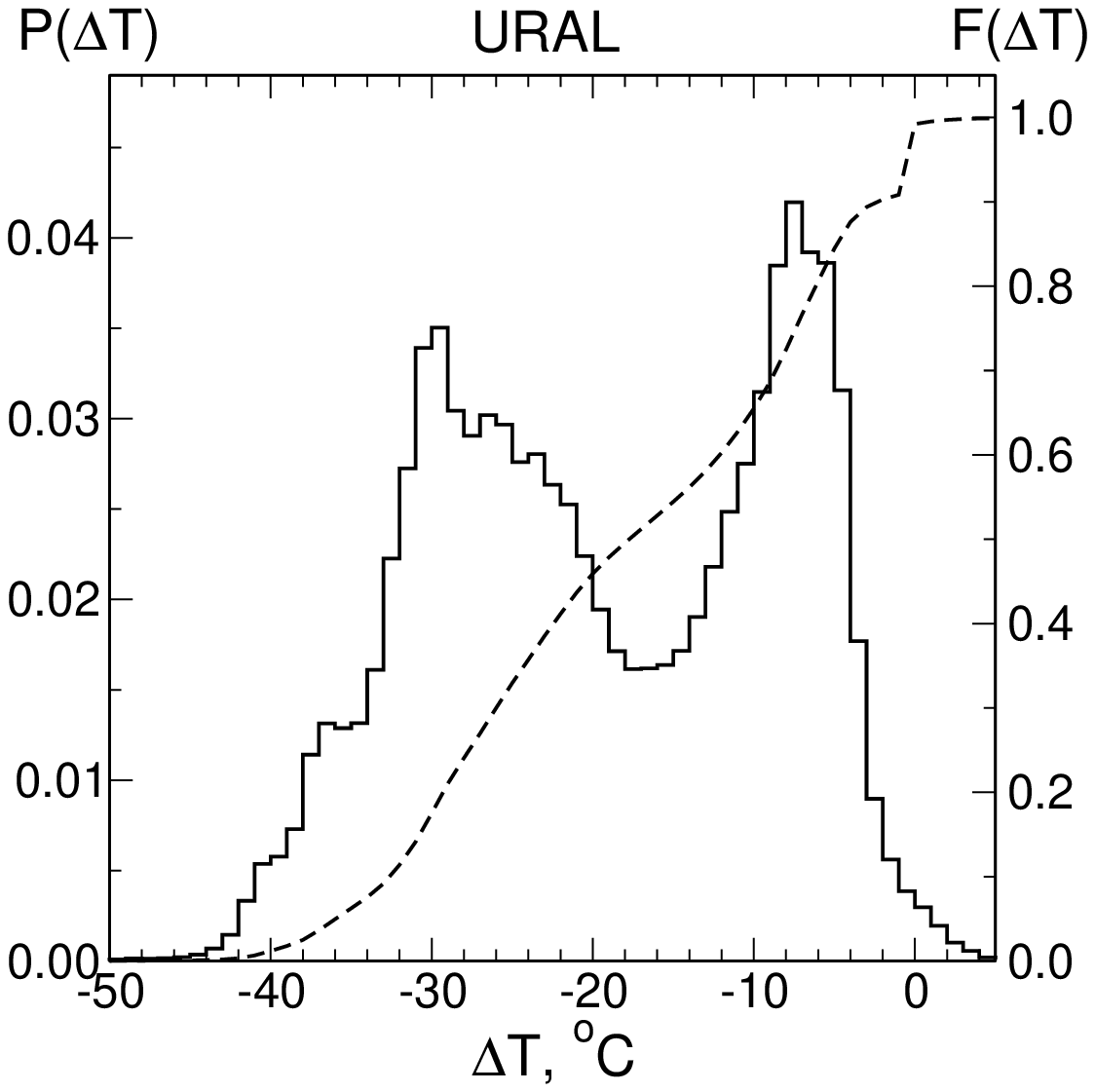,width=6.5cm} &
\psfig{figure=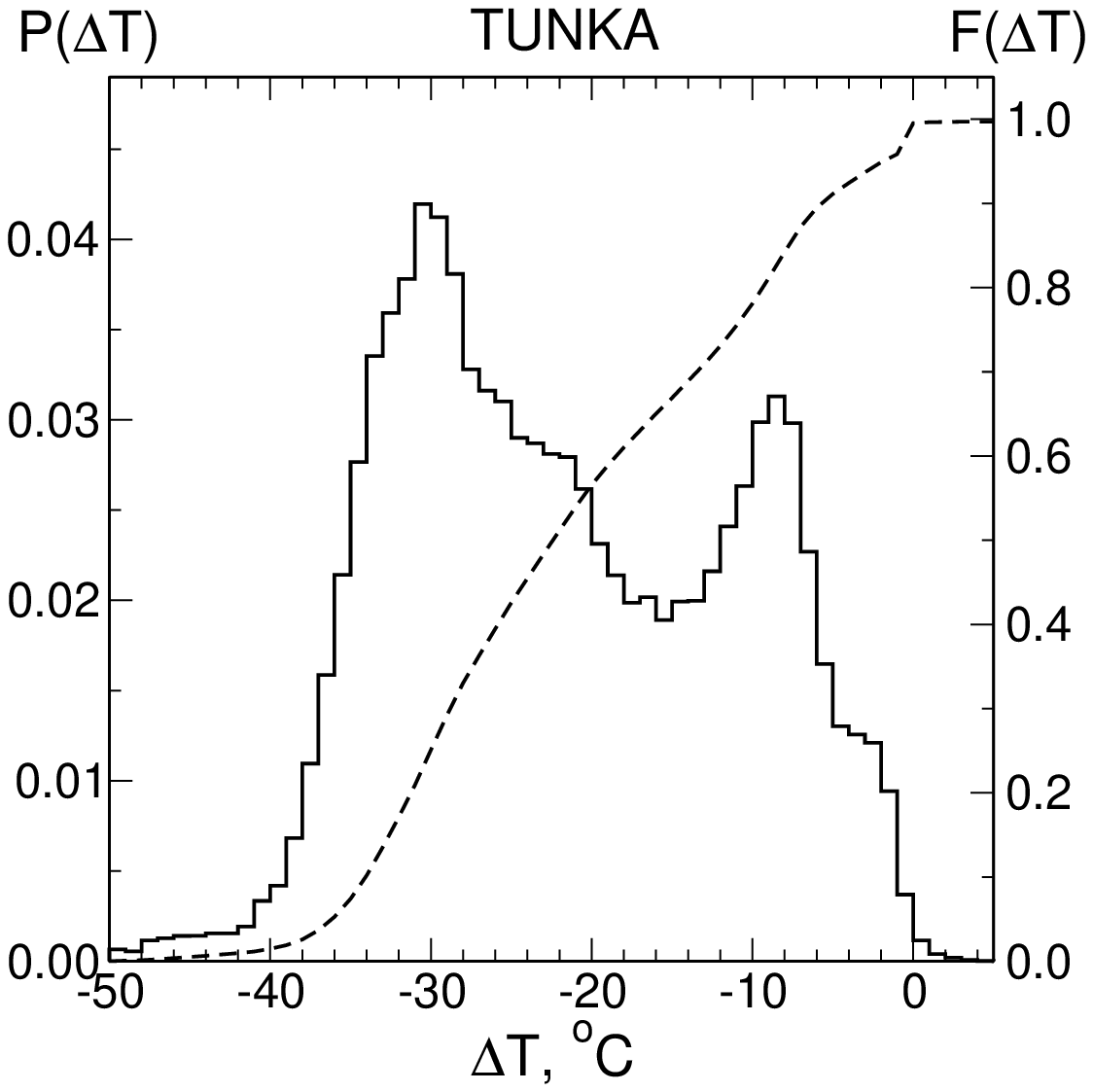,width=6.5cm} \\
\psfig{figure=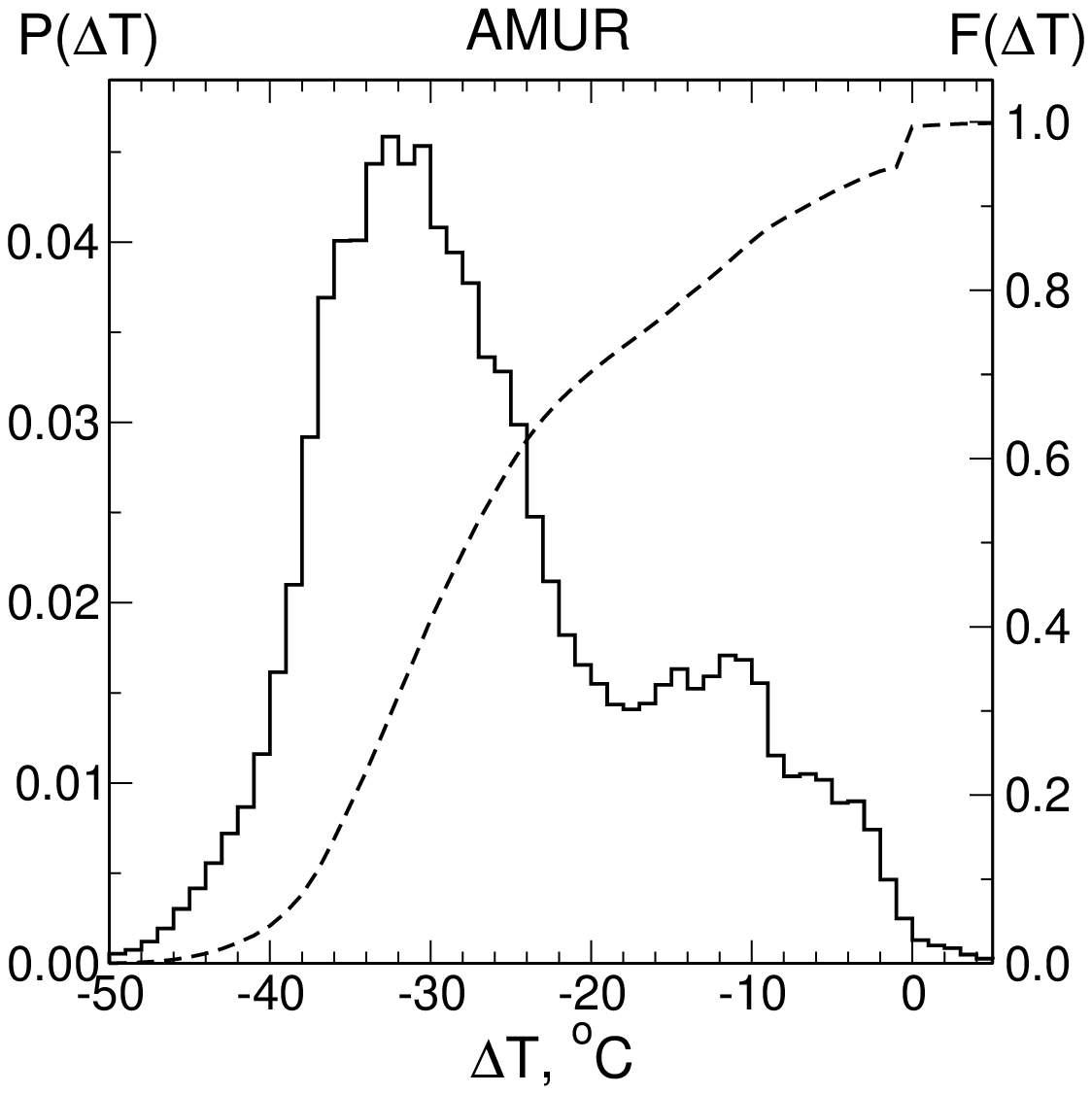,width=6.5cm} &
\psfig{figure=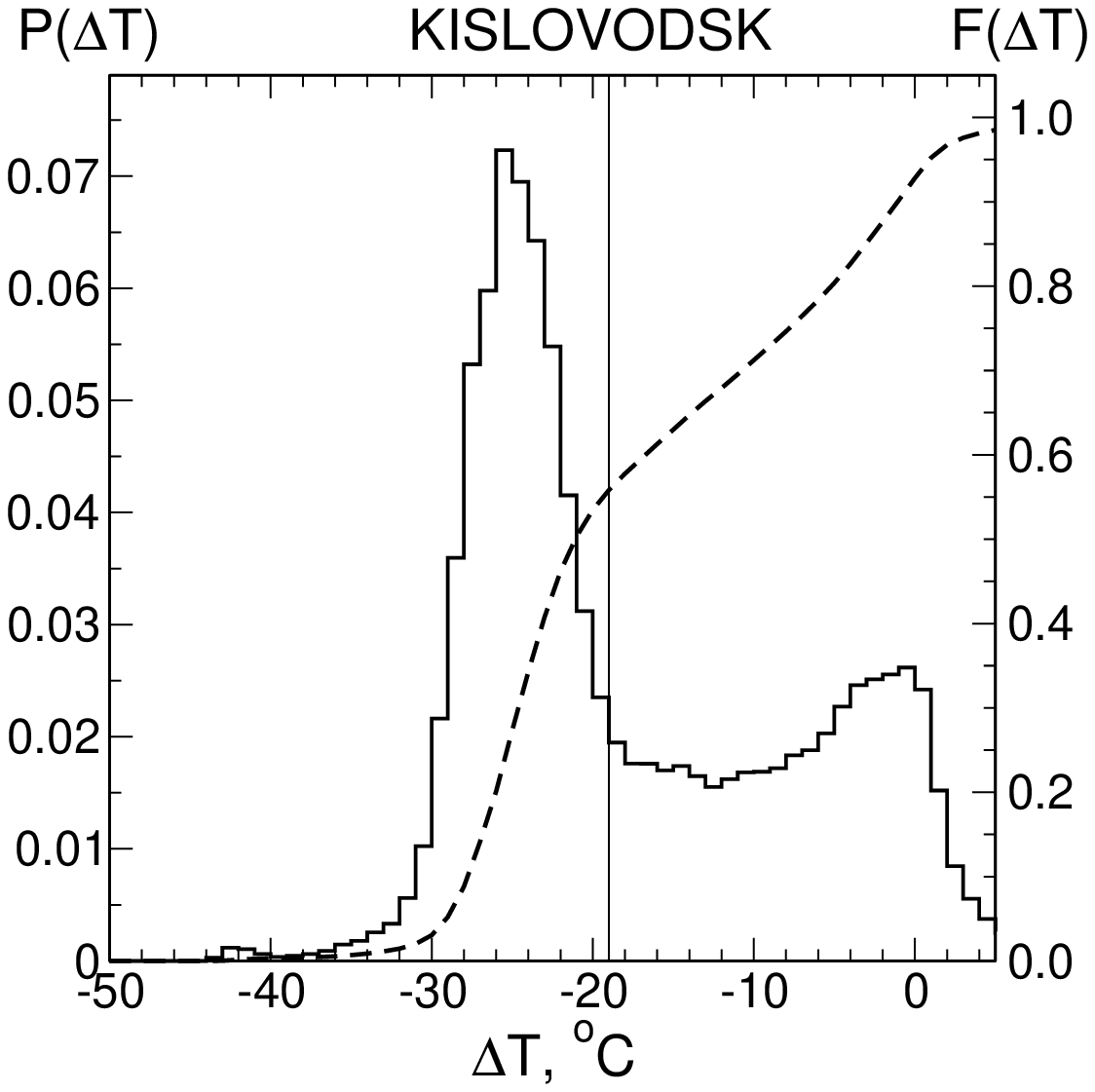,width=6.5cm} \\
\end{tabular}
\caption{The sky temperature distribution for four nodes; the clear sky criterion threshold is shown in bottom right diagram. Step-like distributions are differential; the dashed line are cumulative one (see left- and right-hand axes, respectively
\label{fig:time_clear}}
\end{figure}

It is evident that the longitudinal extent of the network is yet far from full and there is still a large fraction of potential alerts happening in day-time for all the network nodes. The time span (see Fig.~\ref{fig:time_coverage}) from $3^h$ to $9^h$ UT in the winter and $0^h$ to $14^h$ UT in the summer is unavailable for pointings. Yearly averaged amount of this span is $9.5^h$ which corresponds to the 60\% temporal coverage. Therefore, further expansion of the network over longitude is very desirable. Currently, steps are being taken to install a new site on the Canary Islands, Spain. Entering the Tenerife node into operation will increase the coverage up to 70\%.

As follow from Fig.~\ref{fig:time_coverage}, the nearly full (100\%) coverage of time is possible installing the MASTER node at the location with eastern longitude about $6^h$ in the southern hemisphere (latitude $ -35\div -40^\circ$) which is near the Conception city of Chile. Taking into account the clear time fraction of the MASTER nodes and the usual less than 50\% hemisphere availability in each site, this would boost the overall probability of the arbitrary GRB observation from current 0.2 to slightly less than 0.4.

The clear night time fraction is essential site selection parameter which was initially collected from various literature and other sources. After 1 -- 3 years passed in test and routine observations, the considerable amount of weather statistics was collected at sites. The sky clearness parameter used is the brightness temperature of the sky minus ambient temperature. This measure is obtained by the infrared Boltwood cloud sensor\footnote{http://www.cyanogen.com/cloud\_main.php} working in the 10 micron bandpass. This method of determination of the sky availability for observations is widely used at other observatories as well \cite{sanchez2008,calibrBS}.

The temperature differences for 4 network sites are shown in Fig.~\ref{fig:time_clear}. The differential distribution maximums at low temperatures correspond to practically clear skies. The threshold which makes a dissection of cloudy weather from observational one is set to $-21^\circ$C for all sites except for the highland one. At Kislovodsk, the peak is much more pronounced and narrower so the threshold is safely shifted to $-19^\circ$C. It should be stressed that this clear sky definition includes not only the so called photometric conditions but also partly cloudy weather which is suited for alert observations.

Table~\ref{tab:1} presents the observation condition parameters including the yearly astronomical nights duration (sun altitude below $-18^\circ$), total yearly amount of clear night sky time and the clear skies percentage during the astronomical night. The Teide observatory parameters are taken from the paper \cite{casiana}, data for Kislovodsk site are taken from \cite{kgo2010}.

It is clear that the outcome of the accompanying observational programs of the MASTER project -- the sky survey, supernovae search etc -- is mainly driven by the total clear sky time. Current value of the astronomically useful time is $\approx 7000$ hours per year for all the operational sites which will increase soon up to 9400 hours.

The same table provides the median value of the clear night ambient temperature $T_{med}$. This parameter determines both the observational conditions via absolute humidity and the severity of robustness demands to the hardware in use.

The night sky background brightness (in $V$ magnitudes per sq. arcsec) is taken from \cite{cnzano2001} while for the La Palma island observatory it comes from \cite{ORMsky1998}.

\begin{table}
\caption{The total night time duration $FNT$ and clear night time duration $CNT$ for the MASTER network sites.\label{tab:1}}
\smallskip
\centering
\begin{tabular}{lrrrrr}
\hline\hline
Net node \ups &  $FNT$, h     & Доля  $CNT$    & $CNT$, h & $T_{med},{}^\circ$C & Background \\[5pt]
\hline
MASTER-URAL \ups  & 2659  & 46\% & 1223 & $-5.5$  &  21.2--20.5 \\
MASTER-TUNKA      & 2984  & 57\% & 1700 & $-6.5$  &  22.0--21.9 \\
MASTER-AMUR       & 3068  & 70\% & 2146 & $-10.5$ &  20.5--19.5 \\
MASTER-KISLOVODSK & 3291  & 57\% & 1876 & $+1.8$  &  21.9--21.7 \\[5pt]
\hline
MASTER-CANARIES\ups & 3534  & $\sim$70\% & 2474  & $\sim +5$ & 21.9--21.7  \\
\hline\hline
\end{tabular}
\end{table}

\section{MASTER II instrumentation}
\label{equip}

Each observatory was aimed to be equipped with as much as possible standardized hardware configuration in order to provide 1) the more homogeneous data set and 2) the minimal expenses for the soft- and hardware maintenance. The principal element is the {\em twin catadioptric fast telescope} installed on the equatorial mounting at both sides of the mount head.

\begin{figure}
\centering
\psfig{figure=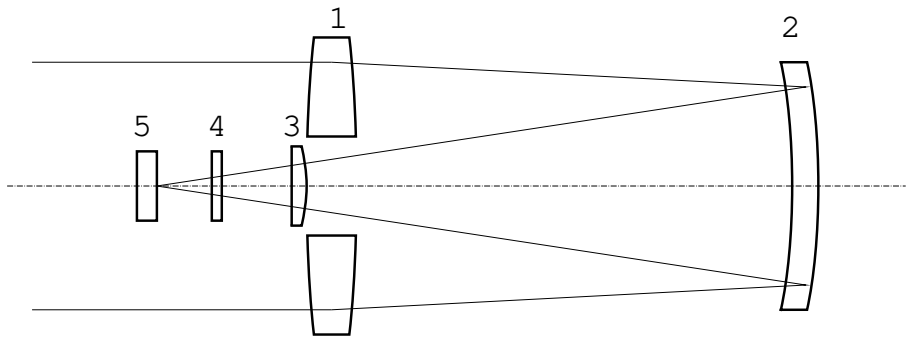,width=12cm}
\caption{Simplified sketch of the Hamilton optical system. 1 -- positive entrance aperture lens, 2 -- Mangin mirror; 3 -- field lens; 4 -- optical filter; 5 -- CCD detector.
\label{fig:hamilton}}
\end{figure}

The Hamilton optical system of the telescopes contains a slow positive lens at the entrance pupil and primary Mangin mirror at the bottom of the tube (Fig.~\ref{fig:hamilton}). The focal plane is in front of the lens so the backfocal distance of  $\approx 100$~mm is provided for the filters and detector assembly. The field flattener lens is located near the entrance aperture and is used for fine focusing having the movement range $\pm 10$~mm.

The system is optimized taking the glass filter thickness of 5~mm. The telescope primary characteristics are following:
\begin{itemize}
\item[---] Entrance aperture diameter $D=400$~mm
\item[---] Focal ratio $F/2.5$;
\item[---] Effective focal length $F=1000$ mm;
\item[---] Plate scale $s=206''/mm$;
\item[---] Maximal field of view diameter $2\omega=4^\circ$;
\item[---] Central obscuration (by the CCD camera body) is $q=25$\% of area;
\item[---] Unobstructed aperture area is $S=940 \mbox{ cm}^2$.
\end{itemize}

The working field of view is restricted by the detector in use to $2.1^\circ\times2.1^\circ$ which is considerably lower than the maximal available field. So, the used camera Apogee Alta~U16M with the detector format $4K\times4K$ shows good images everywhere across its area. This CCD provides the scale of $1.85^{\prime\prime}$/pixel.

The tubes are denoted as ``eastern'' and ``western'' for convenience meaning their telescope parking position allocation. Their supports provide two modes of tubes collimation: 1) the collinear position (both optical axes coincide with the mount pointing vector) and 2) the diverged position (optical axes are at $\pm 1^\circ$ with respect to the pointing vector). The first mode is used for transient alert observations simultaneously in two photometric bands or linear polarization states, while the diverged mode is normally used in surveys.

\begin{figure}
\centering
\begin{tabular}{cc}
\psfig{figure=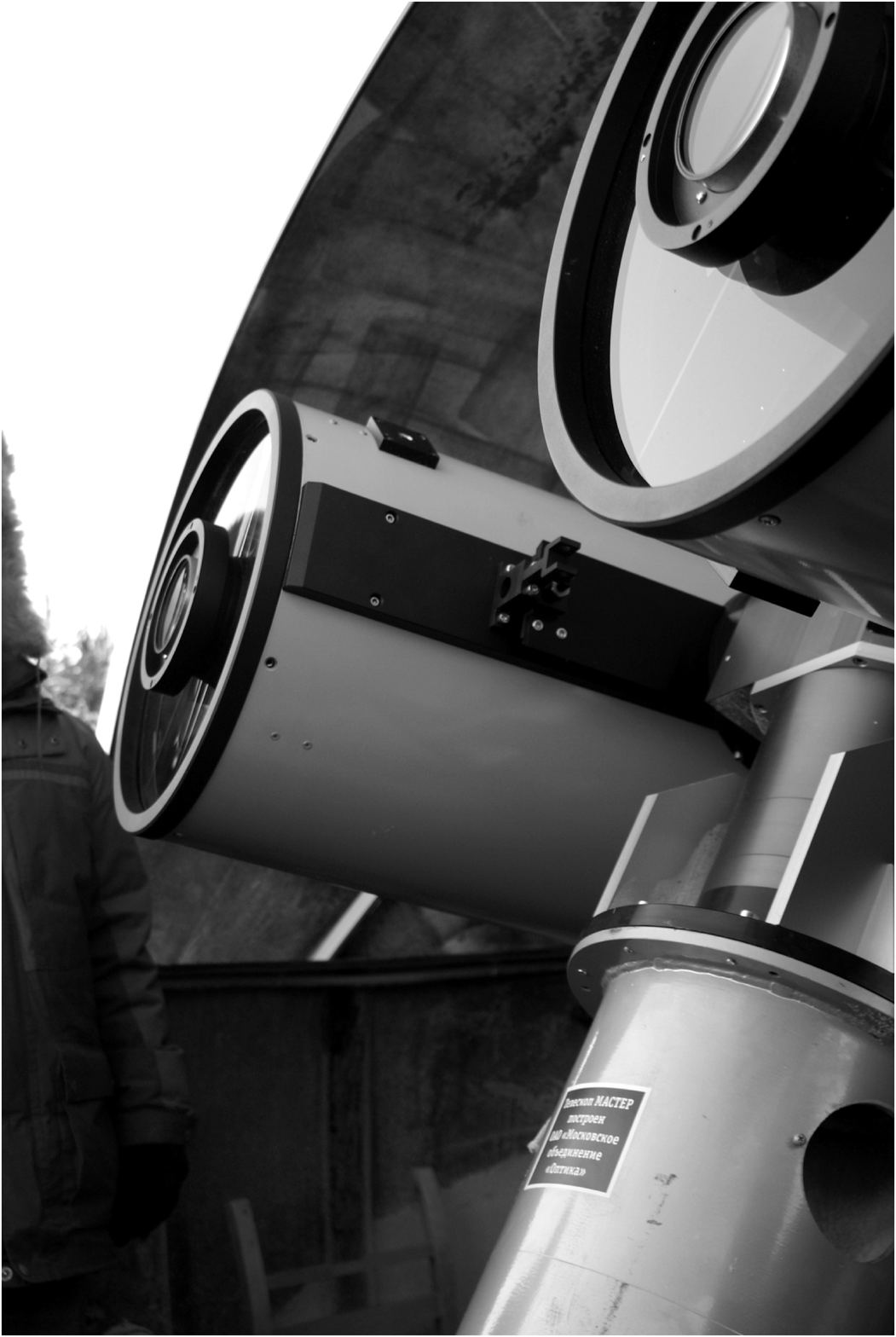,height=7cm}&
\psfig{figure=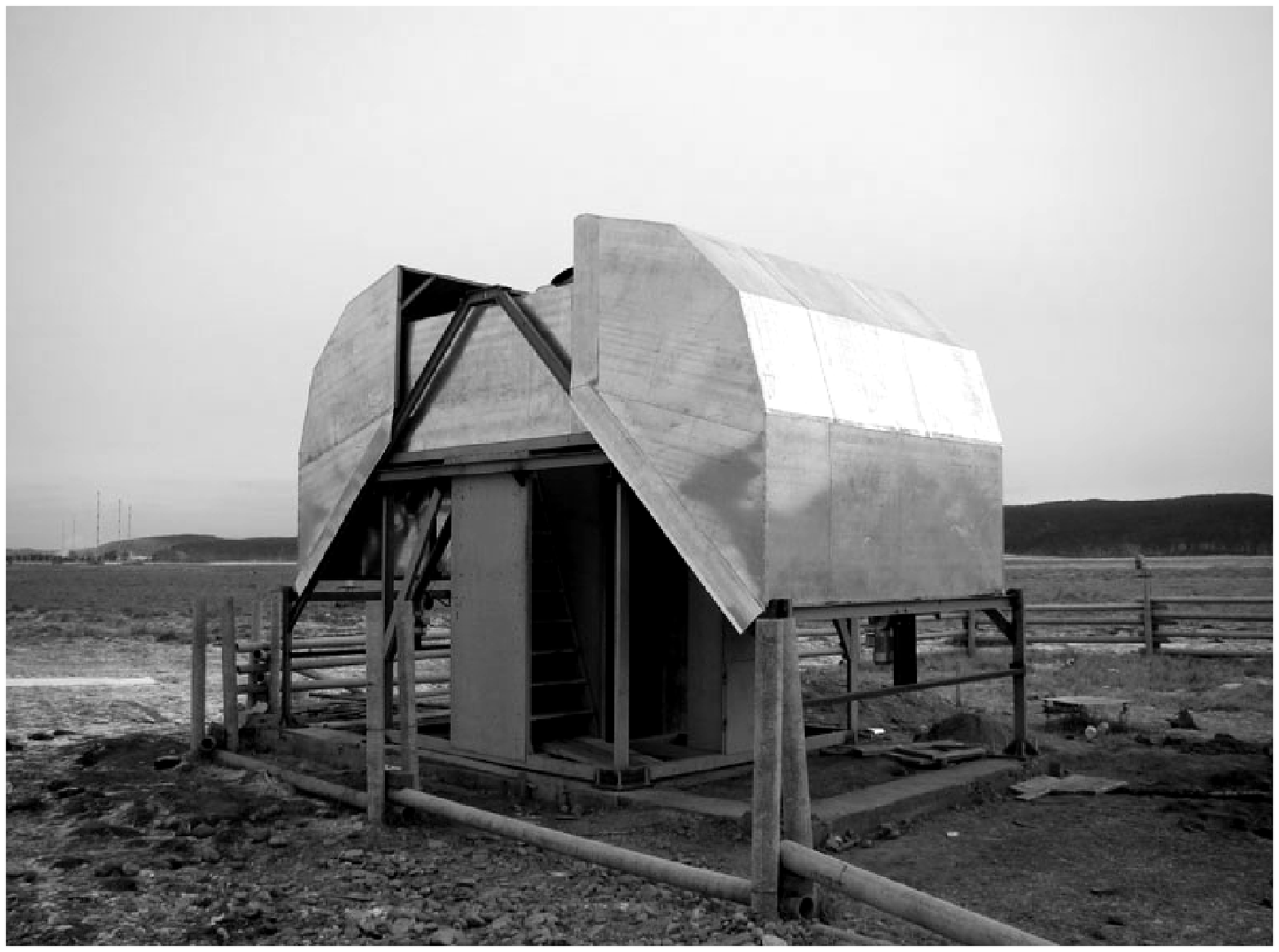,height=7cm}\\
\end{tabular}
\caption{Left: MASTER~II telescope assembly at Kourovskaya astronomical observatory, with no photometric units and CCD attached. The NTM~500 mount is installed on a special console inclined at the latitude angle of the site. Right: custom steel enclosure for Irkutsk university testing area in Tunka valley.  The twin telescope is mounted on the top of a concrete pier in the center of the enclosure.
\label{fig:ural}}
\end{figure}

Since the optical tube assembly gravity center is near the geometrical one, the conventional German type  equatorial mount does not allow for symmetrical tubes attachment. The special console is manufactured as an inclined extender of the mount which adapts the mount to the North pole elevation at each site. The console is visible in the left panel of Fig.~\ref{fig:ural} below the mount head.

The mount in use for MASTER~II telescopes is the {\em ASTELCO NTM~500 mount} based on the direct drive torque motors. The absence of gears allows for speeds up to $30^\circ$/s and also provides the needed robustness for sometimes extreme working conditions (especially at Siberian sites). Being well aligned by azimuth and elevation and properly calibrated, this mount provides the pointing precision of $\sim 10^{\prime\prime}$. Meanwhile, its declination axis is relatively slim and in windy nights ($\ge 5$~m/s) may develop the vibration of up to $10^{\prime\prime}$ -- $20^{\prime\prime}$ amplitude.

\begin{figure}
\centering
\psfig{figure=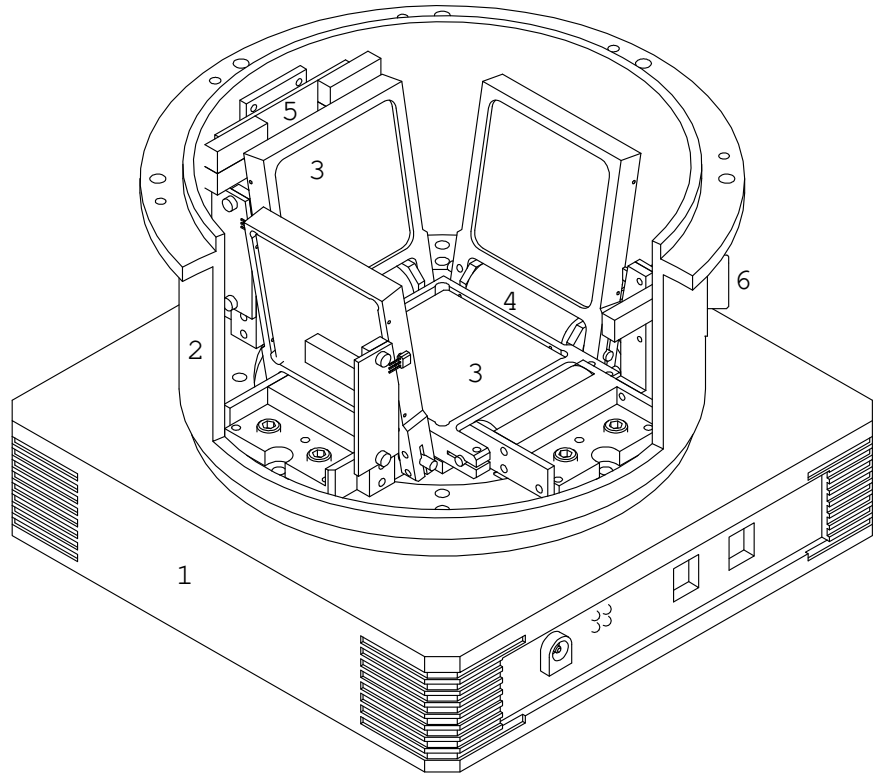,width=12cm}
\caption{The photometric unit structure. 1 -- CCD camera Alta U16M; 2 -- the photometric unit case; 3 -- filters $50\times50$~mm, of which one is on-axis; 4 -- the filter motors; 5 -- control electronic board; 6 -- control and power connector.
\label{fig:photon}}
\end{figure}

The custom made {\em photometric unit} (Fig.~\ref{fig:photon}) provides the four photometric filters or polaroids capability. It serves also as a mechanical interface between the CCD camera and the telescope. The particular set of filters installed in eastern and western telescopes depends on the actual long-term scientific program. Normally both tubes are equipped with Johnson $V$ and $R$ filters for surveys in the diverged mode, one polaroid with the polarization direction perpendicular to the one in the opposite tube assembly, and $B$ or $I$ Johnson filter in the free slot left to provide the simultaneous multicolour photometry in the collinear mode. Polaroid axes are set in two ways with respect to celestial sphere: in MASTER-KISLOVODSK and MASTER-TUNKA sites, the axes are directed at positional angle $0^\circ$ and $90^\circ$, in other two observatories --- at angles $45^\circ$ and $135^\circ$.

In order to protect the assembly from bad weather {\em two types of enclosure} are used (see Fig.~\ref{fig:ural}): the fiberglass clam-shell 3.6-meter Astrohaven dome (MASTER-KISLOVODSK and MASTER-URAL) and steel custom enclosure for more harsh conditions of Siberia (MASTER-TUNKA and MASTER-AMUR). Both enclosures provide the full-hemisphere access except for elevations below $+5^\circ$ where one or both tubes may appear partially obscured. Also, the standard dome controllers of Astrohaven got burnt due to bad insulation and were replaced by the custom made controllers used for all types of enclosures.

As mentioned above, a very important component of an automated system is a {\em cloud sensor}. Apart from the measurement of the sky cloudiness by its brightness temperature, it also provides the additional weather parameters: ambient temperature, precipitations flag and coarse estimate of the wind speed. Currently used is the Boltwood sensor which nevertheless is not absolutely immune to problems.  These devices can be incorrectly calibrated for ambient temperatures below $-20^\circ$C; also two of them broke within the first year. The current conditions are used not only as a safety means but also for scheduling of the auxiliary programs of observations.

For making the most important synchronous gamma-ray burst observations, each MASTER~II telescope is equipped with {\em two very wide field cameras}. These are fast frame rate industrial CCD cameras from AVT company (former Prosilica) GE4000 having a detector format of 4008$\times$2672 pixels and an area of $24\times 36$~mm. The readout noise of the VWF cameras is about $30\,e^-$/pixel.

The VWF cameras in Kislovodsk site are equipped with Canon EOS EF II 85/1.2L lenses having apertures 70~mm and focal length 85~mm. Siberian sites Tunka and Amur employ Zeiss 85/1.4 ZF lenses having a bit smaller apertures with the same focal length. These optics provide the angular fields of $24^\circ\times16^\circ$ with the plate scale $21^{\prime\prime}\mbox{/pixel}$. Note, that such a scale results in a background of $\approx 14^m$/pixel in a moonless night.

Cameras are attached as ``piggy-back'' devices to the main telescopes and pointed with a slight overlap in such a way that they monitor the maximal area around the main field of view of the telescope.

The same cameras\footnote{http://www.alliedvisiontec.com/uploads/media/Prosilica\_Camera\_News\_-\_Issue\_10\_-\_April\_2009.pdf} are used in a paired standalone VWF installation located in Kislovodsk observatory, one near the solar station and another near the 2.5-m SAI telescope building \cite{Ada2010a,Ada2010b}.

Finally, two simple web-cams are making permanent surveillance of the under-dome space and outdoors area for safety reasons.

\section{The MASTER telescopes capabilities}

The potential of the MASTER telescopes is defined by many factors, both by design and by observation conditions. In order to assess the importance of various factors, let us estimate the detected signal strength given the commonly used energy relations. The light from the 20-th $V$-magnitude class A0\,V star arrives to the Earth atmosphere at the rate of 8.2 photons/s per the telescope aperture. Given the average atmospheric transmission of 0.75 in this band, the characteristic optics losses of 0.25 and quantum efficiency of the KAF16803 CCD detector near the 5000\AA\ equal to 0.55, the equivalent signal becomes $\sim 2.5\,e^-$/s. So, the handy estimate is $\sim 1\,e^-$/s for the 21-th magnitude star.

Since the plate scale is $1.85^{\prime\prime}$ per the resolution element (CCD pixel), then the background sky signal equals the signal from the object with $V_{sky}-1^m\!\!.34$. Using the night sky brightness data in moonless nights for different MASTER sites (Table~\ref{tab:1}) one can derive that the typical sky background in images in the $V$~band varies from $1.4\,e^-$/s to $9\,e^-$/s and is even more in the $R$-band.

\begin{figure}
\centering
\psfig{figure=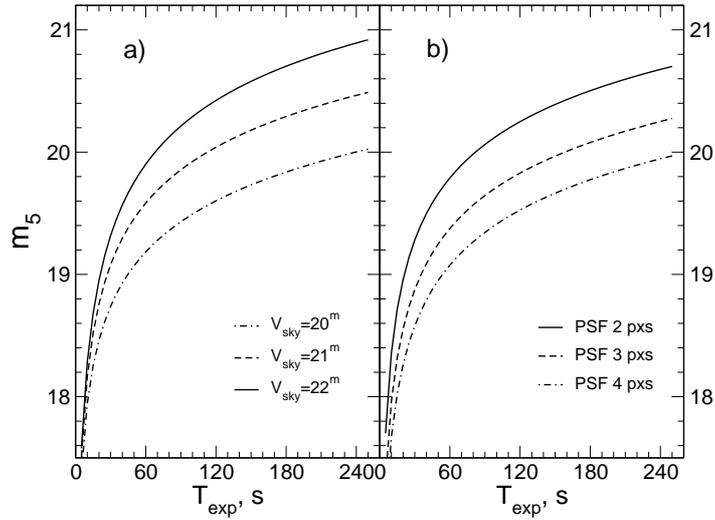,width=8cm,angle=-90}
\caption{The estimate of the limiting $V$-magnitudes ($S/N=5$) as a function of the exposure duration $T_{exp}$: a) given different night sky background b) given varying telescope focusing. The typical CCD readout noise of  $8\,e^-/px$ is assumed.
\label{fig:limits}}
\end{figure}

The measurement of the dark current in the cameras in use revealed that the datasheet value of $0.1 e^-$/s/pixel at the $-30^\circ$C detector temperature is an upper limit, the real dark current is twice as low. In such circumstances, the night sky background dominates the dark current and there is no sense in cooling the cameras deeper. The temperature setpoint stability is much more important than its particular value. The reason is hot pixels of moderate generation rate which are not masked and in case of the temperature fluctuations may generate the false detections.

The typical readout noise of the Apogee Alta U16M camera is $7-9\,e^-$/pixel and is equivalent to the background signal of $50-100\,e^-$/pixel. Such a signal is accumulated from the sky after less than a minute. Hence, the readout noise is a limiting factor at such short exposures while longer exposures are already limited by the sky. This is illustrated by the computed curves of the detection limits shown in Fig.~\ref{fig:limits}. It is clear that at 60~s exposures the dependence character is changed.

The leftmost graph shows the relations for different night sky levels given the focusing is close to ideal. As expected, the night sky brightness variation by $2^m$ causes the detection to shift by $1^m$, while the computed detection limit constrained by $S/N = 5$ is $20^m\!\!.6$. The graph to the right presents the curves illustrating the importance of precise focusing: the images having the doubled FWHM are characterized by the detection limit which is worse by $0^m\!\!.7$.

The real images analysis obtained in different conditions supports the conclusions drawn above. The typical 180~s exposure taken at dark sky ($\approx 20^m\!\!.5$) provides the mean magnitude of weakest measurable stars of $20^m\!\!.0$.

\section{Digital hardware structure of the MASTER~II node}

\label{infohard}

\begin{figure}
\centering
\psfig{figure=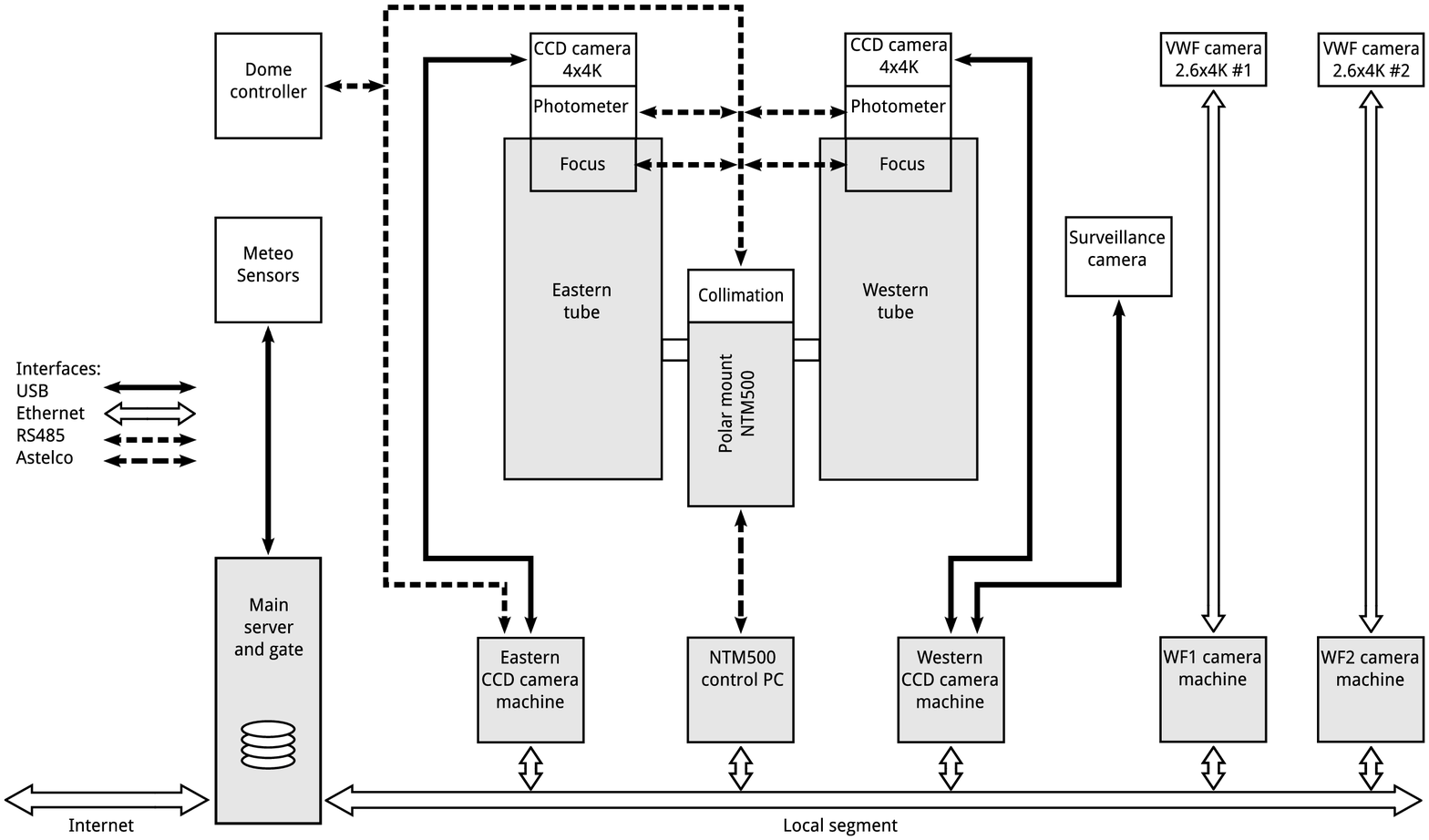,width=20cm,angle=90}
\caption{The general structure of the MASTER~II telescope complex. The various interface types are shown.
\label{fig:struct}}
\end{figure}

The digital hardware structure of the MASTER II complex reflects its functional aiming. It is optimized for effective physical and lower logical levels operation for data and command transfer for the robotized telescope and astronomical and auxiliary equipment control. The overall scheme of this hardware connections is shown in Fig.~\ref{fig:struct}.

The complex is organized as a distributed computer system for data acquisition and processing which is united with help of local Ethernet segment. The core of the complex is a server based on the two-processor 8-core platform Intel Xeon 54xx equipped with a high performance RAID-system. It provides the following functions:
\begin{itemize}
\item[---] The gateway of the complex linking the local net to Internet using VSAT equipment.
\item[---] The precise time service based on {\em ntpd} making the synchronization with Internet time servers and/or a local GPS receiver. Accuracy of the synchronization is about 3~ms and in a worst  case is better than 25~ms.
\item[---] Meteo data supply: the local weather station additionally equipped with an infrared cloud sensor is attached to the server.
\item[---] The database PostgreSQL server is used for storage of meta-information about the saved astronomical images, objects detected on them and auxiliary data.
\item[---] HTTP-server service for external access to observation control and to image and object databases.
The web-interface is protected with SSL encryption and the operators authentication and access right assignment.
\item[---] Data storage: a dedicated disk subsystem stores the original images as well as the processing results in the FITS format. The images are loss-less compressed and stored in an observation date-ordered structure of directories.
\item[---] The GCN-center alerts  program: a permanent socket connection is maintained with the GCN server for prompt delivery of the gamma-ray events observed by space observatories.
\item[---] The observations supervision with help of the continuously running software.
\item[---] Observation scheduler on the basis of pre-selected observation strategies and lists of targets introduced by operators.
\item[---] The image processing system providing automatic processing of acquired images.
\end{itemize}

All the hardware responsible for the safety operation is powered from uninterrupted power supply (UPS) with the capacity providing the normal system shutdown in case of a power cut.

The following interface types are used for communicating with the commercial devices:
\begin{itemize}
\item[---] CCD cameras Alta 16M control, under-dome survey cameras, cloud sensor --- USB\,2.0;
\item[---] VWF cameras --- Gigabit Ethernet (GigE);
\item[---] Communication of the NTM~500 mount computer with the mount itself is made via dedicated interface of ASTELCO company.
\end{itemize}
The variety of physical interfaces is encapsulated on the system logical level (device drivers) while the inherent logical features of them are handled at the low level of software.

The control of the custom-made optical tube assembly mechanisms and enclosure hardware is performed via the RS485 physical interface using custom made controllers. This way of command and data communication is well protected against interfering signals and is galvanically decoupled from the main computer which connects to the line via special RS232/RS485 converters. The simple custom logical protocol of communication is well tested during more than 10 years in various astronomical equipments.

The MASTER-KISLOVODSK observatory is somehow special since it additionally includes two standalone VWF installations located some 800~m from each other \cite{Ada2010b}. Each installation consists of two machines with attached VWF cameras and a compact computer running a mount and an enclosure. Also, the main server room is more than 200~m away from the telescope. So, the local net of the site MASTER-KISLOVODSK incorporates both copper and optical fiber lines interconnected via media-converters. On the other hand, the MASTER-URAL site has no VWF subsystem shown in Fig.~\ref{fig:struct}.

\section{Software network structure}
\label{infosoft}

All the machines of the complex are running the GNU/Linux operation system which is best suited for local, distributed and remote work. The distribution of the hardware is naturally mimicked by the distributed character of the software.

The software consists of several independently running components which are communicating to each other when necessary. This simplifies the logical structure of the components and enhances the stability and flexibility in the system organization. Such a way, part of functionality is provided by the operation system itself, including the parallel execution of different programs in course of online data processing.

\begin{figure}
\centering
\psfig{figure=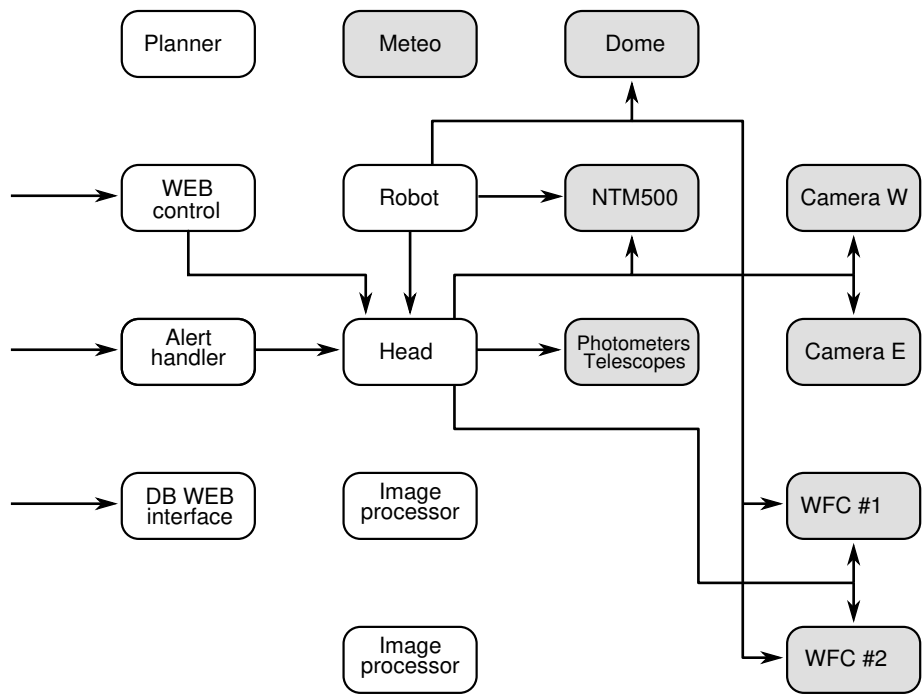,width=12cm}
\caption{The general scheme and network connections of software components of MASTER~II telescope. The arrows point from clients to server programs. The components serving hardware are shown in grey color.
\label{fig:psockets}}
\end{figure}

Each program component provides the single major logical task or function. The optimal division of the observation process on such relatively independent parts is essential here. The priority is given to the task itself, not to the service of a particular device although in many cases the task is indeed directly related to given devices. The basic inter-program communication method is the network TCP/IC connections where some programs perform the role of servers and others work as clients.

The general software structure and mutual relations are given in Fig.~\ref{fig:psockets}. Grey coloured modules work on the periphery machines and serve the particular devices. It should be stressed that client-server architecture implies the use of a dedicated logical command protocol. Such a protocol\footnote{See http://curl.sai.msu.ru/mass/download/doc/sv\_ug.pdf} was developed earlier for another project of Sternberg institute and is based on enumerated command messages with a restricted set of command words followed by optional parameters. The essential feature of this protocol is the obligatory acknowledgement commands received, parsed and executed. The advantage of this protocol is an ease of manual work via any text console which is necessary for system debugging and while working in non-standard situations.

\begin{figure}[b]
\centering
\psfig{figure=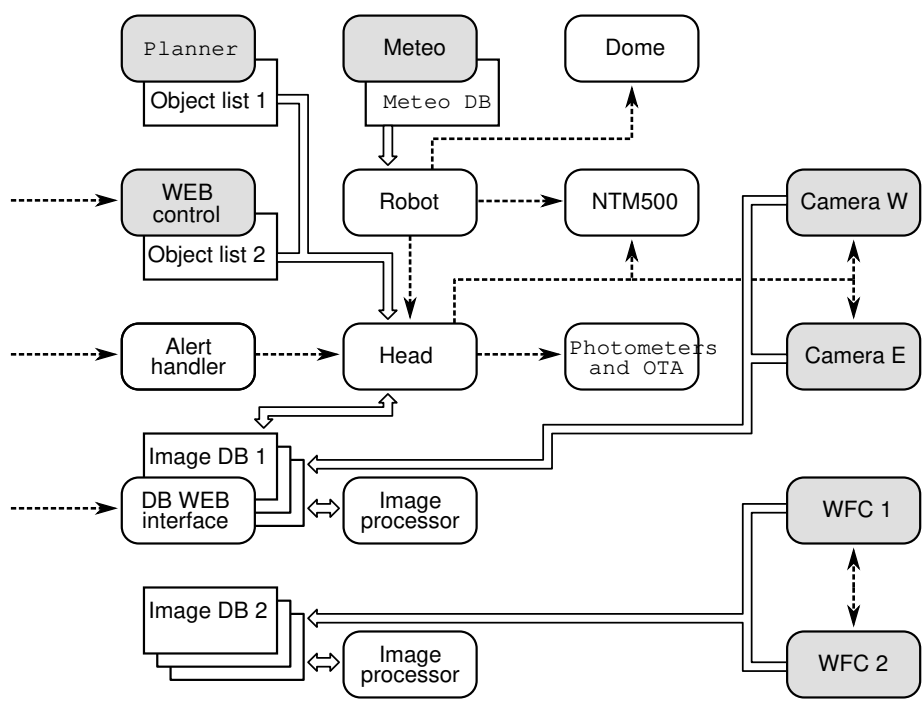,width=12cm}
\caption{Logic structure of the command and data transmission in the MASTER~II software. Dashed arrows -- commands, double line arrows -- data transmission. Programs which generate data for databases are shown in grey.
\label{fig:data_flow}}
\end{figure}

The raw and processed data transmission (mainly images) is implemented using database requests. Three main databases (DB) maintained in the system are the image DB, the weather DB and the object DB (the list of targets for observations). In Fig.~\ref{fig:data_flow} the main relations between the program components and DBs are shown. Auxiliary tools (perl and shell-scripts) involved in data transmission to the server and storing them in the DB are not shown.

The VWF cameras store their frames in a separate DB~2 which is also accessed via a dedicated web-interface. This subsystem is served by a separate image processor due to the specifics of the large image area and high image acquisition rate. These images are kept in the database for no more than 2~days.

\section{Observations scheduling and making}
\label{sec:shed}

In course of observations, the synchronization of the work of separate software components is done by a special program {\sl Head} started on the server. After the {\em initialization}, the program connects to the needed  components as a client and executes the observation stages via commands or data requests. The particular sequence of pointings and acquisitions is made given the data retrieved from the objects list which is  formed by the scheduler or manually, using the respective web-interface. Three modes of observations are supported: {\em survey}, {\em special} and {\em alert} observations.

The {\sl Head} is not responsible for checking of the observation conditions that is performed by a separate daemon {\sl Robot} which task is a continuous monitoring of weather conditions (clear night) from evening to morning twilights. It is {\sl Robot} which makes the decision about observations {\em possibility}. Just as the decision is made, {\sl Robot} connects to the enclosure program in order to open it and to the {\sl Head} to initialize it. Likewise, if conditions become worse, {\sl Robot} parks the {\sl Head} program and closes the dome. Additional responsibility of the daemon is starting and stopping the surveys and checking the telescopes and domes state after the dawn in order to ensure the equipment safety. Also, {\sl Robot} monitors the UPS state and cancels observations (first of all closing the dome) in case the power cut is detected.

The survey starting/stopping is made so special in terms of the programs assignment because of the fact that weather conditions may be not suitable for survey or searching works while alert events are responded in any conditions which are safe for equipment. This case corresponds to the third {\sl Head} state -- the standby state.

The information on gamma-ray bursts (alert events) which is received from the Gamma Center Network \cite{GCN} server of NASA\footnote{http://gcn.gsfc.nasa.gov/} by a special daemon-server {\sl Alert handler}, is immediately handed over to the {\sl Head} by a special connection. This request will be ignored in a parked state only when observations are impossible. In the standalone or survey observations state this alert request will be immediately processed even if the current exposure will have to be aborted.

The alert observations logic is following: 1) point a telescope to the GRB position, 2) set the needed filters and correct the focus respectively, 3) put the tubes in the collinear position and 4) start the exposure with the duration $\sim 0.2\,\Delta t_{GRB}$, where the $\Delta t_{GRB}$ is the time passed since the GRB moment. The observations thus started are continued as long as possible or until stopped manually. A special case when another alert arrives while observing the first one is rare but handled in a special way: the current exposure is completed (not aborted) and afterwards the {\sl Head} starts to serve the new event.

The Fermi alerts \cite{fermia} are observed during 5 minutes. This is a trade-off due to their coarse coordinates which makes their shooting quite a lucky case given the small field of view but the pointing is indeed performed in a hope to capture them with the VWF cameras.

Survey observations are performed in different ways depending on the survey task. The main survey is executed for the search of the non-catalogued objects of various nature and maintaining the base of reference images in the MASTER telescopes photometric systems. This is made in the $R$-band or white light with the diverged tubes capturing the maximal field in a single exposure. The supernovae search survey is a bit more special. It is made with collinear tubes and usually with $V$ and $R$ filters installed.

The object priority of the database {\sl Object list \#2} filled in manually via the web-interface is the highest one. In case it contains no objects suitable for current time, the pointing coordinates are retrieved from the database {\sl Object list \#1} which is generated by the {\sl Planner} program.  Element of the survey is an ``area'' which is a sky region sized by the telescope FOV and referenced by its centre coordinates. The tactics of surveys is to image the given area several times separated by different time spans, from minutes to hours.

The {\sl Planner} program takes the surveys priority into account and corrects it by individual telescope factors, current astronomical and weather conditions. It maintains the given tactics filling in the {\sl Object list \#1} by actual objects or areas. For next observation, an object is selected which has a highest weight computed from a number of parameters. These parameters include equatorial and horizontal coordinates, distances from the Galactic plane, ecliptic, Moon or last observed area and from the SWIFT spacecraft current FOV position, a number of archive frames of this area already collected, the moment of previous observation of the area, a number of galaxies in the area and a number of known Ia-type supernovae.

In the process of survey, the {\sl Head} program performs the quality control of the acquired images. If necessary, the exposure is repeated. This function also tracks the telescope focusing and, if necessary, starts the auto-focusing procedure.

Every night, the program {\sl Head} starts a process of acquisition of needed calibration images, arranged as a separate observational task. Dark frames (for different exposure times) and bias frames are obtained in the evening and morning twilight when Sun is below $-1^\circ$. If the sky is cloudless, the flat field images are shot with a 1~s exposure in the zenith in all photometric bands while Sun has an altitude between $-5^\circ$ and  $-7^\circ$. Then, over-exposed and under-exposed flats are discarded.

The VWF cameras are started by the {\sl Robot} daemon after which they continuously acquire images during the night. The supplementary information for their FITS headers is provided by the {\sl Head} program.

\section{Automatic image processing and databases}

It is worth noting here that the database on images is the multi-table one: the first table keeps the raw images, including calibration data, the second corresponds to the primarily processed images and finally there are tables containing subimages cropped around automatically classified objects of different kinds.

The image processing is started after its insertion in the database ($\approx 25$~s after the image readout which are needed for image transmission to the server and insertion itself). The alert images are separated from survey frames from the very beginning. Alert images have the highest priority for processing. If necessary, the image processing programs may be started in parallel so up to~8 (the number of CPU cores) images may be reduced simultaneously.

The first stage is the primary data reduction including the bias and dark frame subtraction and flat-fielding. Calibration master frames are prepared in advance as a library of images for all the detector temperature setpoints used as described above. The calibration images are stored in the same table as the reduced science frames for subsequent inspection if needed.

The next stage is the objects detection and their coordinate and photometric referencing. As a base for object detection and classification, the {\sl SExtractor} package \cite{sex} developed for the TERAPIX project was adopted and additionally tuned for our images application. The table of detected objects is used for the sky patch identification and calculation of the coordinates conversion coefficients. The coordinates are taken from the Tycho~2 catalogue which contains the precise coordinates of majority of stars up to $11^m\,.5$ \cite{Tycho}. In case the reference stars are lacking, the USNO~B1.0 \cite{usnob} stars are added.

The photometric reduction is made using the same catalogues given the photometric band in use. It is performed using minimization of differences between the instrumental and catalogue magnitudes for all the reference stars selected on the current frame. If a catalogue magnitude in the band is not accessible, a simple transformation is calculated and used. The precision of such a calibration is $\approx0^m.2$ which is quite enough for exploration tasks and transients detection. On the same stage, the image quality averaged across the frame (FWHM and ellipticity) and the limiting magnitude are estimated. Finally, a record  is added in the database  table of the processed images.

The detected objects are classified into three categories:
\begin{enumerate}
\item the known star --- the object identified using its coordinates (with the allowed difference not more than $3^{\prime\prime}$ with catalogue position) and fitting the catalogue by magnitude ($\pm0.5^m$);
\item ``flare'' --- the object on the place of catalogue one but having higher difference in magnitude ({\em both} negative and positive);
\item unknown -- the object absent in catalogues.
\end{enumerate}
The known stars are excluded from subsequent consideration and these are only class 2 and 3 -- the potential transients (PT) --- which are analyzed. The image artifacts may also pretend being PT: unaccounted hot pixels, diffraction spikes from bright stars, on-edge objects. A special program filter is marking these cases. Then the PT is checked for a coincidence with additional catalogues (AGN, asteroids and SNe).

All the PT objects (including marked as potentially false) are added to the database in the object table and the subsequent analysis is made already using this table, not the image. The table already contains the information on all the previous area observations. These data allow for further PT classification.

The {\em supernova} candidates are selected from unmarked PT which are located within the doubled radius of the closest galaxy 25-th isophote taken from the HYPERLEDA catalogue \cite{Hyper}. Then the previous observations are checked. Unknown objects lying within more than 0.75 fraction of the 25-th isophote radius are accounted as candidate ``pure'' supernova \cite{masha,2009nr} and inspected with a first priority. All the SN candidates are additionally cleaned from asteroids casually projecting onto galaxies.

The {\em optical transients} (OT) are selected only from objects observed more than twice. As candidate OTs, the unmarked PTs are considered having a large ($>2^m$) magnitude deviation for which an archive frame exists showing no object on the detection place. Such a frame must be obtained at least one week before the observation and its limiting magnitude should allow for the candidate OT detection.

The selection of candidate {\em asteroids} is made solely in repeated area observations. The first observation allows only addition of the known minor planets to the database if they are identified in the asteroid orbit database \cite{Lowell}. While searching for asteroids, a typical trajectory is assumed and PTs of previous observations are considered as a candidate asteroid if they lie along the same trajectory.

It is worth to note that the PostgreSQL database system in use is a free object relational database management system which allows to make all the described requests, searches and comparisons in the real-time mode.

Each frame inserted in the image database may be retrieved either in FITS format for subsequent thorough analysis or as a jpeg picture for visual inspection in complex cases. It is possible to retrieve a full image copy or a particular region. Also, the database contains the averaged parameters of the image quality (mean FWHM and ellipticities of stellar images, detection limits etc) and its processing status (successfully, reasonably, error).

\section{MASTER II telescopes external control}

For the MASTER network control, the dedicated web-interfaces are developed which allow to monitor the state and manage the process at any observatory. The main page of each site contains all the necessary information on current conditions: weather parameters, Sun altitude, dome and telescopes state, main program {\sl Head} and {\sl Robot} status. Also, the central regions of last acquired images are shown.

Additional web-interface allows an operator to ease the observation data analysis. The web-control and results analysis systems represent the user interfaces to browse the content and/or modify the data in the databases in use. The flexible system of user authorization discriminates the operators by a number of functions which they are allowed to use for a telescope and data analysis control. These access permissions are assigned and modified by the network administrator.

As already mentioned in Section \ref{sec:shed}, operative management of special programs is carried out through the DB table {\sl Object list \# 2} used by the {\sl Head} program as a priority object list. This table can be updated and edited by an authorized observer at any time using a WEB interface special form where the position of the object, the period of observation, exposure, spectral bands and the task priority are be specified.

It is not needed to coordinate alert mode observation at different observatories because an information on gamma-ray bursts arrives almost simultaneously to all telescopes, and if the external conditions allow these observations, they are performed. Obviously, such an approach increases the probability to successfully capture the event and to get the best data. This is even more important in case of polarization measurements where one needs to eliminate the ambiguity of results obtained with only two polarisation directions available from one telescope observation.

Although the automatic overall control of surveys and special tasks is not yet implemented, its components and a common database are developing. In any case, the observing programs at the observatories are mutually coordinated and approved by the PI of the MASTER network. It is also planned to implement the system  of reciprocal alerts when one of the telescopes detects an interesting transient object.

Special web-pages are designed for each class of objects, whose search is made in real-time (supernovae, optical transients, asteroids). For example, the sky patch is displayed around the potential transient and the full supplementary information on the respective data reduction is shown as well. Additionally, the same sky area images are generated given the most informative DSS and SDSS surveys, the repeated and archive images. For the SN candidates, the known information on the parent galaxy is given as well as the computed distance of the SN candidate from its center.  In case of candidate asteroids detection, the most probable orbit is computed along with the current tangent velocity and output accompanied by the simulated movie illustrating this motion.

\section{Conclusions}

To give a sense of the information accumulation rate, the number of images recorded in different MASTER sites is shown in Table~\ref{tab:2} for March, 2011. It counts only those frames which were processed successfully and which appeared to be suitable for subsequent astronomical use. The data presented show that alert observations occupy less than 10\% of time, all the rest time is spread between the background (surveys) and special observation tasks.

The numbers in Table~\ref{tab:2} depend mainly on elapsed time of the regular observation on those sites. On average, one node of the MASTER network produces about 10--30~GB of astronomical images per night of observation. To date, the total number of obtained frames exceeds 250\,000 and the MASTER II database stores more than 15~TB of astronomical images.

\begin{table}[b]
\caption{The number of acquired images and the accumulated exposure in the MASTER network observatories as for the 2011, mid-March. \label{tab:2}}
\smallskip
\centering
\begin{tabular}{l|rr|rrr}
\hline\hline
\ups             & \multicolumn{2}{c|}{Surveys} & \multicolumn{2}{c}{Alerts} \\
Site              &  $N$  & $E$, min      & $N$ & $E$, min & \\[3pt]
\hline
MASTER-URAL \ups  & 30\,561 & 76\,928     & 1\,941 & 4\,713 &   \\
MASTER-TUNKA      & 27\,688 & 59\,900     & 1\,124 & 2\,465 &   \\
MASTER-AMUR       & 53\,088 & 64\,960     & 3\,696 & 3\,825&   \\
MASTER-KISLOVODSK & 63\,214 & 179\,850    & 6\,131 & 17\,500 &   \\[3pt]
\hline\hline
\end{tabular}
\end{table}

Not all of these frames are of excellent quality, about 7-8 percent have elliptical star images (their axes ratio $>1.35$) due to wind-generated vibrations. Nevertheless, these images can also be used for various kinds of preliminary estimates.

\begin{figure}[t]
\centering
\psfig{figure=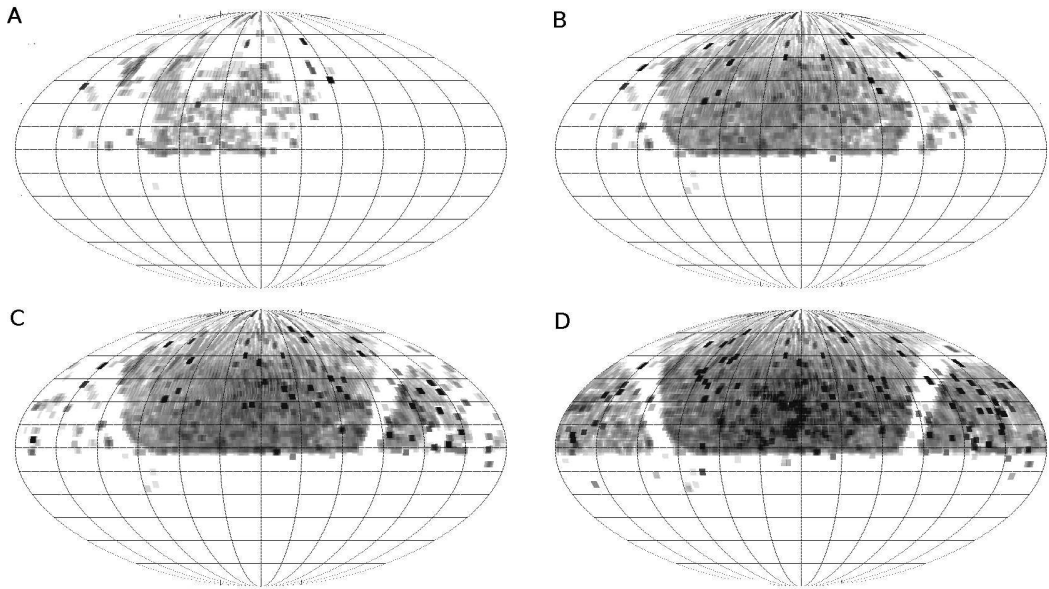,width=14.0cm}
\caption{The map of observations performed at MASTER-TUNKA node: A -- after 30 days from the site start-up, B -- after 90 days, C -- after 180 days and D --- after a year.
\label{fig:survey}}
\end{figure}

In Fig.~\ref{fig:survey}, the visual representation of the survey execution at MASTER-TUNKA node of the MASTER network is shown for the one-year period since summer 2010. The similar picture can be seen at other nodes --- after about half a year since the node deployment about 90\% of the accessible sky is covered by the own archive images. Sky patches shown most dark correspond to most deep study made in course of alert events, possible transients or supernovae observations.

The efficiency of the MASTER~II network is validated by its total input in the GRB optical observations: in the 2009-2010 period about 50 alerts were observed and in 6 cases the optical emission of GRBs was detected. At the present time, the MASTER II network provides more than 50\% of first pointings to Swift GRBs. In the same period, 13 supernovae were independently detected in spite of that the SN search is only an assistant programme.

During the MASTER II network exploitation, some technical and algorithmic problems were diagnosed. For the most part, they were solved. Besides, the experience accumulated in real observations drives us to modify many details both in image acquisition and in image processing. Further efforts to improve the performance of the network and resulting data quality are continuing. However, the correctness of the principal solution to create a distributed network of compact twin telescopes was definitely confirmed.

\begin{acknowledgements}
The MASTER~II project was supported by Russian foundation of basic researches No 09-02-12287-ofim, and by Ministry of Education and Science -- the Innovative programme of Ural university and Contract 02.740.11.0249. The bulk of support for the project came from the Development Programme of Lomonosov Moscow state university. The authors are grateful to authorities of the Moscow state university, Sternberg astronomical institute, Pulkovo observatory, Ural state university, Irkutsk state university and Blagoveshchensk pedagogical university who contributed to the practical implementation of the project. They also thank the staff of observatories involved in the MASTER~II network. The authors express their gratitude to the Director of the Teide observatory for his willingness to host the fifth MASTER network node.
\end{acknowledgements}

\end{document}